\documentclass[twocolumn ,showpacs,preprintnumbers,  
               superscriptaddress]{revtex4-1}
\usepackage{graphicx, fancybox}
\usepackage{amsmath,amssymb}

\begin{document}

\title{On the effect of secondary protons on baryon and proton 
number cumulants in event-by-event analysis}

\author{Hirosato Ono}
\email{ono@kern.phys.sci.osaka-u.ac.jp}
\affiliation{
Department of Physics, Osaka University, Toyonaka, Osaka 560-0043, Japan}

\author{Masayuki Asakawa}
\email{yuki@phys.sci.osaka-u.ac.jp}
\affiliation{
Department of Physics, Osaka University, Toyonaka, Osaka 560-0043, Japan}

\author{Masakiyo Kitazawa}
\email{kitazawa@phys.sci.osaka-u.ac.jp}
\affiliation{
Department of Physics, Osaka University, Toyonaka, Osaka 560-0043, Japan}

\begin{abstract}

We investigate the effects of secondary (knockout) protons,
which constitute about $20\%$ of the observed 
protons at STAR, on the higher order cumulants of proton 
and baryon numbers measured by event-by-event analyses 
in relativistic heavy ion collisions.
We argue that the contribution of this background effect 
on the cumulants is expressed by a simple formula, 
and that hence their effects can be removed in the experimental 
analysis.
It is discussed that this background effect has non-negligible
contribution to recently observed proton number cumulants at 
STAR, especially the third-order one, and that the removal of 
this effect is crucial to investigate the thermodynamical properties
of the primordial hot medium appropriately.

\end{abstract} 
\date{\today}
\maketitle

\section{Introduction}
\label{sec:intr}

The beam energy scan (BES) program at the Relativistic Heavy 
Ion Collider (RHIC) has recently accumulated an abundance of 
experimental data on heavy ion collisions at various 
collision energies per nucleon, $\sqrt{s_{\rm NN}}$ 
\cite{Aggarwal:2010wy,Mohanty:2011nm,STAR:QM2012,PHENIX:QM2012}.
Because hot media created by collisions with 
different $\sqrt{s_{\rm NN}}$ follow different 
trajectories on the temperature ($T$) and baryon chemical 
potential ($\mu_{\rm B}$) plane during their time evolution, 
careful inspection of these data will enable us to 
understand the global phase structure 
of QCD on the $T$-$\mu_{\rm B}$ plane.

Much attention has been paid to fluctuations of conserved 
charges as experimental observables to probe the primordial 
thermal properties of the hot medium \cite{Koch:2008ia}.
Fluctuations are characterized by cumulants.
It is theoretically pointed out that the cumulants 
have characteristic behaviors on the $T$-$\mu_{\rm B}$ 
plane, reflecting the phase structure of QCD 
\cite{Stephanov:1998dy,Asakawa:2000wh,Jeon:2000wg,
Ejiri:2005wq,Stephanov:2008qz,Asakawa:2009aj,
Friman:2011pf,Stephanov:2011pb}.
These behaviors can in principle be measured in heavy 
ion experiments by observing fluctuations in 
event-by-event analyses.
Analyses on higher order cumulants are being carried out in 
the BES program \cite{Aggarwal:2010wy,Mohanty:2011nm,
STAR:QM2012,PHENIX:QM2012,Luo:2011tp}.
Because cumulants of conserved charges can also 
be measured on the lattice \cite{lattice}, 
fluctuations of conserved charges can be used 
to compare these real and virtual experiments 
as well as theoretical predictions.

For conserved charges, the time-evolution of fluctuations 
in a volume $V$ is slow for large $V$, and hence 
fluctuations in the final state reflect those 
generated in the primordial stage when the rapidity 
range of detection is taken sufficiently large 
\cite{Asakawa:2000wh,Jeon:2000wg,Shuryak:2000pd}.
Indeed, the recent experimental result on 
electric charge fluctuation by ALICE collaboration at 
the Large Hadron Collider (LHC) suggests 
that the fluctuation created in the quark-gluon phase 
well survives until the final state at the LHC energy
$\sqrt{s_{\rm NN}}=2.76$ TeV \cite{Abelev:2012pv}.

The $\sqrt{s_{\rm NN}}$ dependence of net-proton number cumulants 
has been analyzed by STAR Collaboration at RHIC last 
two years \cite{Aggarwal:2010wy,Mohanty:2011nm,STAR:QM2012}.
It should be first emphasized that the net-proton number is 
not a conserved charge, and should not be
considered as a proxy of the net-baryon number.
Whereas the results on the proton number cumulants are
sometimes identified and compared with the baryon number
ones in the literature, this identification could lead 
one to a wrong conclusion \cite{KA1,KA2}.
To appropriately compare the experimental results with 
theoretical predictions on baryon number cumulants,
the former has to be translated into the latter in the first
place \cite{KA1,KA2}.

Nevertheless, although with such reservations,
the experimental results on the net-proton 
number cumulants show some interesting features. 
To interpret the experimental results on the cumulants,
they are often compared with the hadron resonance gas 
(HRG) model \cite{HRG}.
If the fluctuations are generated in a thermal medium 
described well by the free hadronic degrees of freedoms, the 
experimental results should well reproduce the HRG model.
Therefore, the HRG model is used as the baseline 
from which the deviation of the cumulants encodes 
information on the non-hadronic and/or non-thermal 
feature of the primordial hot medium.
In the HRG model, the probability distribution of the number
of each baryon in a phase space is Poissonian
(the net number distribution is thus given by the 
Skellam distribution) \cite{HRG}.
In the experimental results on proton number cumulants 
at STAR, small but statistically-significant deviations 
from this baseline are observed \cite{STAR:QM2012}.
These deviations are expected to be due to non-trivial 
properties of the hot medium from the above argument, 
and theoretical and experimental verifications and 
investigations of these deviations are important for 
extracting the global nature of the QCD phase structure 
from these experimental data.

In experiments, various backgrounds affect 
measurements of observables. In the measurement 
of fluctuation observables in event-by-event 
analyses, background effects would make the distributions
of the fluctuations approach Poissionian ones.
The observation of the deviation of the distributions 
from the Poissonian ones thus would be subject to such 
background effects.
In order to see the signals of the non-hadronic 
property of the hot medium encoded as the deviation
from the Poisson distribution, therefore, careful 
treatment on the background effect is necessary.

In the present study, we focus on the background effect 
of secondary (knockout) protons, which are protons 
knocked out from materials in the detector by 
reactions with particles created in the hot medium.
We derive relations to remove this background effect
from higher order cumulants, and 
argue that this background effect has a non-negligible 
contribution to the experimental measurement of the 
cumulants of net-proton and net-baryon numbers.

\section{secondary protons}

\subsection{General properties}
\label{sec:2nd}

In the present study, we consider the effects of secondary 
protons on the cumulants of the net-proton and net-baryon 
numbers in relativistic heavy ion collisions.
Secondary protons are protons which come from not 
fireballs created by heavy-ion collisions,
but materials in the detector.
Protons in the detector can be knocked out by 
reactions with particles from the fireballs, 
and enter the detector.
A major part of the secondary protons can be identified 
and eliminated in the experimental analysis by introducing 
a cut for the distance between the reconstructed trajectory 
of the proton and the collision point, which is usually 
represented by $d_{\rm ca}$ in the literature \cite{Abelev:2008ab}.
Some of them, however, have reconstructed trajectories 
which pass near the collision point, and are 
misidentified as true protons created in the fireballs.
The abundance of these misidentified protons for 
various transverse momentum ($p_T$) bins estimated by an 
event generator for STAR for $62$ GeV AuAu collision is 
shown in Table~I in Ref.~\cite{Abelev:2008ab}.
With this table and the $p_T$-dependence of the proton 
yield, it is estimated that about $20\%$ of protons 
detected in this experiment are the secondary ones for 
the typical $p_T$ range used in the event-by-event 
analysis of proton number cumulants, $0.4<p_T<0.8$GeV.
In the following, we denote the number of misidentified
secondary protons in each event as $N_p^{\rm 2nd}$.
Notice that the anti-proton number is not affected
by this background effect, because materials in the 
detector do not contain anti-particles.

The background effect by secondary protons is usually 
subtracted from experimental results on {\it average} 
values using estimates with event generators.
Their effects on the second- and higher-order 
{\it cumulants}, however, are nontrivial, and to the 
best of the authors' knowledge, this effect has not 
been taken care of in the existing experimental results 
on fluctuation observables.
The purpose of the present study is to quantify 
their effects and make it possible to eliminate them 
in the experimental analyses.

In order to obtain such a relation, 
let us clarify several properties of secondary protons.
First, we assert that the probability distribution of the 
number of secondary protons in each event, 
${\cal P}_{\rm 2nd}(N_p^{\rm 2nd})$,
is Poissonian, ${\cal P}_{\rm 2nd}(N_p^{\rm 2nd}) 
= \lambda^{N_p^{\rm 2nd}} e^{-\lambda}/N_p^{\rm 2nd}!$, to a
good approximation, with 
$\lambda = \langle N_p^{\rm 2nd} \rangle$.
To support this statement, we note that 
the probability $r$ that a particle generated in the 
hot medium creates a misidentified secondary proton
is small, $r\ll1$.
In fact, whereas particles of order $10^3$ are created
in each relativistic heavy ion collision, the 
value of $\langle N_p^{\rm 2nd} \rangle$ is of order $10$
\cite{Abelev:2008ab}.
This shows that $r$ is of order $10^{-2}$, which is 
significantly smaller than unity.
Moreover, events where 
secondary protons are created are uncorrelated with
each other, because the knockout 
reactions take place independently in the bulk detector.
These two properties are sufficient to guarantee the 
Possionian nature of ${\cal P}_{\rm2nd}(N_p^{\rm 2nd})$.
Whereas the total multiplicity for each event is not 
fixed even after fixing the collision geometry, and 
the probability to create a secondary proton depends 
on particle species, the Poissonian nature is not 
altered by these features.
The Poisson distribution is specified by a single
parameter $\lambda = \langle N_p^{\rm 2nd} \rangle$, 
and all cumulants take a unique value,
\begin{align}
\langle (\delta N_p^{\rm 2nd})^n \rangle_c
= \langle N_p^{\rm 2nd} \rangle ,
\label{eq:P}
\end{align}
where $\langle (\delta N_p^{\rm 2nd})^n \rangle_c$ is 
the $n$-th order cumulant of the misidentified 
secondary proton number.

Next, let us consider the correlation between 
$N_p^{\rm 2nd}$ and the genuine proton number 
emitted from the hot medium, $N_p^{\rm col}$.
Because of the creation mechanism of secondary 
protons, $N_p^{\rm 2nd}$ for each event has a strong 
positive correlation with the total multiplicity $N^{\rm tot}$.
Since the total multiplicity $N^{\rm tot}$, which is 
dominated by pions, is correlated with $N_p^{\rm col}$, 
in general, $N_p^{\rm 2nd}$ and $N_p^{\rm col}$ are also 
correlated with each other.
On the other hand, when collision events with a fixed 
$N^{\rm tot}$ are concerned, provided that $r$ does 
not depend on particle species or momentum, 
the distribution of $N_p^{\rm 2nd}$, which is Poissonian,
is completely determined and hence has no correlation
with $N_p^{\rm col}$ or any other particle numbers 
emitted from the hot medium.
In experiments, the number of charged particles, 
$N_{\rm ch}$, is used as a proxy of $N^{\rm tot}$, and 
analyses are performed for collision events with 
finite ranges of $N_{\rm ch}$.
For example, $N_{\rm ch}$ is used as the centrality 
selection at STAR, and central to peripheral
collisions are usually classified 
centrality bins with the size of 
$5$ or $10\%$ \cite{Abelev:2012pv}.
In the following, we assume that 
the cut of $N_{\rm ch}$ in the experimental analysis 
is sufficiently narrow to suppress correlations of 
$N_p^{\rm 2nd}$ with $N_p^{\rm col}$, as well as with 
the numbers of anti-protons, baryons, and anti-baryons.

In the analysis of cumulants in STAR experiments, 
cumulants have been estimated by the weighted method 
\cite{Luo:2011tp}. In this method, cumulants are first 
determined with narrow ranges of $N_{\rm ch}$, and 
then they are averaged with a certain wider range of 
$N_{\rm ch}$.
In this method, the correlation between 
$N_p^{\rm 2nd}$ with $N_p^{\rm col}$ vanishes almost 
completely when the value of $\langle N_p^{\rm 2nd}\rangle$ 
determined for each range of $N_{\rm ch}$ is used.

\subsection{Proton number cumulants}

Neglecting the correlation between 
$N_p^{\rm col}$ and $N_p^{\rm 2nd}$, one can write 
the probability distribution of the total proton 
number detected by the detector, 
$N_p^{\rm exp}=N_p^{\rm col}+N_p^{\rm 2nd}$, as 
\begin{align}
{\cal P}_{\rm exp}(N_p^{\rm exp})
= \sum_{N_p^{\rm col},N_p^{\rm 2nd}}
& {\cal P}_{\rm col}(N_p^{\rm col})
{\cal P}_{\rm 2nd}(N_p^{\rm 2nd})
\nonumber
\\ 
&\times \delta_{N_p^{\rm exp},~N_p^{\rm col}+N_p^{\rm 2nd}}.
\label{eq:e2rb}
\end{align} 
Then the moment and cumulant generating functions of 
$N_p^{\rm exp}$ are given by 
\begin{align}
G_{\rm exp}( \theta ) 
= \sum_{N}
{\cal P}_{\rm exp} (N) e^{N \theta}
= G_{\rm col}(\theta) G_{\rm 2nd}(\theta),
\label{eq:G}
\end{align}
and 
\begin{align}
K_{\rm exp}( \theta ) 
= \log G_{\rm exp} ( \theta )
= K_{\rm col}(\theta) + K_{\rm 2nd}(\theta),
\label{eq:K}
\end{align}
respectively, where the generating functions for 
$N_p^{\rm col}$ and $N_p^{\rm 2nd}$ are defined, respectively, as 
\begin{align}
G_{\rm col}(\theta) &= \sum_N {\cal P}_{\rm col}(N) e^{N\theta} 
= \exp{K_{\rm col}(\theta)},
\\
G_{\rm 2nd}(\theta) &= \sum_N {\cal P}_{\rm 2nd}(N) e^{N\theta}
= \exp{K_{\rm 2nd}(\theta)}.
\end{align}
In Eq.~(\ref{eq:K}), $K_{\rm exp}(\theta)$ is given by 
the sum of the two independent generating functions
$K_{\rm col}(\theta)$ and $K_{\rm 2nd}(\theta)$,
owing to the independence of the two probabilities
${\cal P}_{\rm col}(N)$ and ${\cal P}_{\rm 2nd}(N)$.
Cumulants $\langle (\delta N)^n \rangle_c$ are defined by derivatives of the cumulant 
generating function as 
\begin{align}
\langle (\delta N)^n \rangle_c
= \left. \frac{ \partial^n K }{ \partial \theta^n }\right|_{\theta=0}.
\end{align}
Cumulants up to fourth order satisfy 
$\langle \delta N \rangle_c = \langle N \rangle$,
$\langle (\delta N)^2 \rangle_c = \langle (\delta N)^2 \rangle$,
$\langle (\delta N)^3 \rangle_c = \langle (\delta N)^3 \rangle$, 
and $\langle (\delta N)^4 \rangle_c 
= \langle (\delta N)^4 \rangle - 3 \langle (\delta N)^2 \rangle^2$.
From Eq.~(\ref{eq:K}), it is clear that the cumulant of 
total proton number is given by the sum of the two cumulants,
\begin{align}
\langle (\delta N_p^{\rm exp})^n \rangle_c
&= \langle (\delta N_p^{\rm col})^n \rangle_c + 
\langle (\delta N_p^{\rm 2nd})^n \rangle_c
\nonumber \\
&= \langle (\delta N_p^{\rm col})^n \rangle_c + \langle N_p^{\rm 2nd} \rangle,
\label{eq:exp-2nd}
\end{align}
where in the last equality we have used Eq.~(\ref{eq:P}).

Eq.~(\ref{eq:exp-2nd}) relates the experimentally 
observed proton number cumulants, 
$\langle (\delta N_p^{\rm exp})^n \rangle_c$,
with the genuine proton number cumulants in the 
final state in heavy ion collisions, 
$\langle (\delta N_p^{\rm col})^n \rangle_c$.
With the value of $\langle N_p^{\rm 2nd} \rangle$ 
estimated in experiments, this relation 
enables one to eliminate the effect of secondary 
protons from higher-order cumulants.

\subsection{Net-proton and baryon number cumulants}

Next, we apply a similar argument to the cumulants of 
net-proton and net-baryon numbers, $N_{p-\bar{p}}^{\rm col}$
and $N_{\rm B-\bar{B}}^{\rm col}$, respectively.
With the same argument addressed in Sec.~\ref{sec:2nd},
one can regard that the correlations between $N_p^{\rm 2nd}$ 
and the number of each baryon is well suppressed 
with a sufficiently narrow $N_{\rm ch}$ bin.
Assuming the absence of the correlation 
in a set of collision events analyzed, 
the probability distribution to observe $N_p^{\rm exp}$ 
protons and $N_{\bar p}$ anti-protons in the experiment 
is written as 
\begin{align}
{\cal G}_{\rm exp}(N_p^{\rm exp},N_{\bar p})
= \sum_{N_p^{\rm col},N_{\bar p},N_p^{\rm 2nd}}
&
{\cal G}_{\rm  col}(N_p^{\rm col},N_{\bar p})
{\cal P}_{\rm 2nd}(N_p^{\rm 2nd})
\nonumber
\\ 
&\times \delta_{N_p^{\rm exp},~N_p^{\rm col}+N_p^{\rm 2nd}},
\label{eq:Gnet}
\end{align}
where ${\cal G}_{\rm  col}(N_p^{\rm col},N_{\bar p})$
is the probability distribution function that the final 
state of each collision event has $N_p^{\rm col}$ protons 
and $N_{\bar p}$ anti-protons
in the phase space covered by the detector.
Note that the anti-proton number $N_{\bar p}$ does not
receive modification from secondary protons, and 
the distinction between ``exp'' and ``col'' is
not needed.
This relation immediately leads to 
\begin{align}
\langle (\delta N_{p-\bar{p}}^{\rm exp})^n \rangle_c
= \langle (\delta N_{p-\bar{p}}^{\rm  col})^n \rangle_c 
+ \langle N_p^{\rm 2nd} \rangle.
\label{eq:Np}
\end{align}

In Refs.~\cite{KA1,KA2}, formulas to obtain net-baryon 
number cumulants using experimentally-observed proton 
number fluctuations were derived.
These formulas are based on the fact that the 
probability to observe a baryon coming from the hot 
medium as a proton or a neutron in the detector is even 
and uncorrelated for each baryon to a good approximation.
This property leads to 
\begin{align}
{\cal G}_{\rm  col}( N_p,N_{\bar{p}})
=& \sum_{ N_{\rm B},N_{\bar{\rm B}}}
{\cal F}_{\rm col}(N_{\rm B},N_{\bar{\rm B}}) 
\nonumber \\
&\times B_{1/2}(N_p;N_{\rm B}) B_{1/2}(N_{\bar p};N_{\bar{\rm B}}),
\label{eq:GF}
\end{align}
where ${\cal F}_{\rm  col}(N_{\rm B},N_{\bar{\rm B}})$
is the probability distribution to have 
$N_{\rm B}$ baryons and $N_{\bar{\rm B}}$
anti-baryons in the phase space in the final state, 
with the binomial distribution function 
\begin{align}
B_r(N;M) 
= r^{N} (1-r)^{M-N} \frac{M!}{N! (M-N)! }.
\end{align}
Neglecting the correlation between (anti-)baryon 
number and $N_p^{\rm 2nd}$, one can substitute 
Eq.~(\ref{eq:GF}) into Eq.~(\ref{eq:Gnet}), obtaining
\begin{align}
\langle (\delta N_{\rm B-\bar{B}}^{\rm exp})^n \rangle_c^{\rm (rec)}
= \langle (\delta N_{\rm B-\bar{B}}^{\rm col})^n \rangle_c
+2 \langle N_p^{\rm 2nd} \rangle,
\label{eq:NB}
\end{align}
where $\langle (\delta N_{\rm B-\bar{B}}^{\rm exp})^n \rangle_c^{\rm (rec)}$
is the $n$-th order cumulant of the net-baryon number reconstructed
from experimentally-observed proton number fluctuations 
using Eqs.~(9)-(12) in Ref.~\cite{KA1},
while $\langle (\delta N_{\rm B-\bar{B}}^{\rm col})^n \rangle_c$
is the genuine net-baryon number cumulants in the final state.

By employing Eqs.~(\ref{eq:Np}) and (\ref{eq:NB}), one can remove
the effects of the secondary protons from the cumulants of net-proton and 
net-baryon numbers, respectively.

\subsection{Effects on ratios of cumulants}

Next, let us consider how the recent experimental 
results on the net-proton number cumulants are 
modified with the effect of secondary protons 
using Eq.~(\ref{eq:Np}).

Because each cumulant is an extensive quantity 
and proportional to the volume of the system,
experimental results on proton number cumulants 
are usually discussed in terms of their ratios
in order to cancel out the volume dependence.
Eq.~(\ref{eq:Np}) shows that the effect of secondary
protons modifies these ratios as 
\begin{align}
R^{\rm col}_{nm} \equiv
\frac{\langle (\delta N_{ p-\bar{p}}^{\rm col})^n \rangle_c }
{\langle (\delta N_{ p-\bar{p}}^{\rm col})^m \rangle_c }
&=
\frac{\langle (\delta N_{p-\bar{p}}^{\rm exp})^n \rangle_c 
- \langle N_p^{\rm 2nd} \rangle }
{\langle (\delta N_{p-\bar{p}}^{\rm exp})^m \rangle_c 
- \langle N_p^{\rm 2nd} \rangle }.
\label{eq:ratio}
\end{align}
This relation shows that when the ratio is unity,
i.e. the Poissonian value, the effect of secondary 
protons does not alter the ratio.
When the ratio 
$R^{\rm exp}_{nm} \equiv 
\langle (\delta N_{ p-\bar{p}}^{\rm exp})^n \rangle_c /
\langle (\delta N_{ p-\bar{p}}^{\rm exp})^m \rangle_c $
has a deviation from unity, however, 
removing the effect of secondary protons makes 
the difference large,
\begin{align}
|1-R^{\rm col}_{nm}|>|1-R^{\rm exp}_{nm}|.
\end{align}
Substituting $\langle N_p^{\rm 2nd} \rangle 
\simeq 0.2 \langle N_p^{\rm exp} \rangle$ in 
Eq.~(\ref{eq:ratio}), one has
\begin{align}
R^{\rm col}_{nm}
&\simeq 
\frac{\langle (\delta N_{p-\bar{p}}^{\rm exp})^n \rangle_c 
- 0.2 \langle N_p^{\rm exp} \rangle }
{\langle (\delta N_{p-\bar{p}}^{\rm exp})^m \rangle_c 
- 0.2 \langle N_p^{\rm exp} \rangle }
= \frac{ R^{\rm exp}_{nm} - r }{ 1 - r }
\\
& 
= R^{\rm exp}_{nm} -r ( 1 - R^{\rm exp}_{nm} ) + O(r^2),
\label{eq:ratio2}
\end{align}
with $r = 0.2 \langle N_p^{\rm exp} \rangle / 
\langle (\delta N_{p-\bar{p}}^{\rm exp})^m \rangle $.
In the last line, $r$ is regarded small and 
higher order terms in $r$ are neglected;
in the HRG model, $r$ approaches $0.1$ for 
large $\sqrt{s_{\rm NN}}$, while 
$r\simeq0.2$ for $\sqrt{s_{\rm NN}}\lesssim 20$ GeV
for even $m$.

At STAR experiment,
$R^{\rm exp}_{32}\simeq 0.2$ 
at $\sqrt{s_{\rm NN}}\simeq200$ GeV \cite{STAR:QM2012}.
Substituting this value into Eq.~(\ref{eq:ratio2}), one has
\begin{align}
R^{\rm col}_{32} \simeq R^{\rm exp}_{32} - 0.1,
\end{align}
with $r \simeq 0.1$.
The third order cumulant at large $\sqrt{s_{\rm NN}}$ 
thus will receive about $50\%$ modification from the 
effect of secondary protons.

The values of $R^{\rm exp}_{42} $ at STAR, on the other hand,
is slightly smaller than but not far from unity for all 
$\sqrt{s_{\rm NN}}$ analyzed in the BES program \cite{STAR:QM2012}; 
$R^{\rm exp}_{42} \simeq 0.8$ at $\sqrt{s_{\rm NN}}\simeq20$ GeV 
is the smallest value.
With this value of $R^{\rm exp}_{42}$, the difference 
between $R^{\rm exp}_{42}$ and $R^{\rm col}_{42}$ expected 
from Eq.~(\ref{eq:ratio2}) is a few percent and is 
not significant.
While the correction is not large, we note that the 
important signal is in the difference 
from the Poissonian value, $1-R^{\rm col}_{42}$.
In terms of this difference, the effect of 
secondary proton gives rise to about $10\%$ 
modification for $R^{\rm exp}_{42} \simeq 0.8$,
\begin{align}
1-R^{\rm col}_{42} \simeq 1.1( 1-R^{\rm exp}_{42} ),
\end{align}
which would not be negligible in the discussion of 
non-hadronic nature of the primordial fireballs.

While we have discussed the effect of secondary protons
on net-proton number cumulants in this subsection,
when one wants to compare the experimental results 
with theoretical predictions, it is important 
to obtain the net-baryon number cumulants \cite{KA1,KA2}.
When the baryon number cumulants are analyzed in 
experiments, the effect of secondary protons should be 
removed with Eq.~(\ref{eq:NB}).

\section{Discussion and Summary}

While we have concentrated on the background effects by 
secondary protons on higher-order cumulants of the 
net-proton and net-baryon numbers in this study,
similar arguments are also applied to other background 
effects.
In particular, when a background effect can be regarded
independent of the baryon numbers in the final state, 
the background effect can be removed in a  
way completely similar to the argument in the present study. 
One of such candidates is the 
effect of particle misidentification.
On the other hand, the effects of efficiency and acceptance 
of the detector are approximately expressed as a binomial 
distribution \cite{KA1,KA2,Bzdak:2012ab}, and different
treatment is required \cite{KA2}.

In the present study, we considered the effects of secondary
protons on higher order cumulants of the net-proton and 
net-baryon numbers measured by event-by-event analyses.
We argued that the number of secondary 
protons in each event follows Poisson distribution, and 
is uncorrelated with the numbers of each baryon emitted 
from fireballs. 
Using these properties, effects of secondary protons 
on net-proton number cumulants were given in a simple 
form, Eq.~(\ref{eq:Np}).
This formula enables one to subtract the background 
effect from the experimental results on the cumulants.
The formula for the net-baryon number cumulants was 
also given by Eq.~(\ref{eq:NB}).

It is found that the effect of secondary protons makes 
the ratios of the cumulants in experimental measurements
closer to unity compared to the ratios with the number of
the genuine protons coming from the hot medium.
Such an effect will particularly modify the value of 
the third-order cumulant at large $\sqrt{s_{\rm NN}}$.
Since the deviation of the ratio from unity carries
information on the non-hadronic and non-thermal
properties of the primordial 
fireballs, appropriate treatment on this effect is 
important to reveal the genuine thermodynamical nature 
of the primordial fireballs.

\end{document}